\def\year{2021}\relax

\documentclass[letterpaper]{article} 
\usepackage{aaai19}  
\usepackage{times}  
\usepackage{helvet} 
\usepackage{courier}  
\usepackage[hyphens]{url}  
\usepackage{graphicx} 
\usepackage{graphicx}  
\usepackage{type1cm}     
\usepackage{graphicx}     
\usepackage{xspace}     
\usepackage{balance}     
\usepackage{multirow}     
\usepackage[font={bf}, tableposition=top]{caption}     
\usepackage{bold-extra}     
\usepackage{siunitx}          
\usepackage[vlined,linesnumbered,ruled,noend]{algorithm2e}     
\usepackage{booktabs}     
\usepackage{microtype}    
\usepackage{units}     
\usepackage{mathtools}     
\usepackage[show]{chato-notes}
\usepackage{relsize}
\usepackage{caption}
\usepackage[]{inputenc}
\usepackage{xfrac}
\RequirePackage{graphicx,color}
\usepackage[font={small}]{subfig} 
\usepackage{pifont}
\usepackage{footnote} 
\usepackage{balance}
\usepackage{hyphenat}

\makesavenoteenv{table}
\usepackage{mathptmx}

\newcommand{\xhdr}[1]{\vspace{1.7mm}\noindent{{\bf #1.}}}

\newcommand{\citet}[1]{\citeauthor{#1} (\citeyear{#1})}

\newcommand{\ourtitle}{Political Polarization in Online News Consumption} 

\author{
  Kiran Garimella\thanks{Research done partly while at EPFL.}\\
  MIT\\
  garimell@mit.edu
\And
 Tim Smith\\
 Mozilla\\
 tdsmith@mozilla.com
\And
 Rebecca Weiss\\
 Mozilla\\
 rweiss@mozilla.com
\And
  Robert West \\
  EPFL\\
  robert.west@epf\/l.ch
}

\begin{document}
\title{\ourtitle\thanks{Paper accepted at ICWSM 2021. Please cite the ICWSM version.}}
\maketitle

\begin{abstract}
Political polarization appears to be on the rise, as measured by voting behavior, general affect towards opposing partisans and their parties, and contents posted and consumed online.
Research over the years has focused on the role of the Web as a driver of polarization.
In order to further our understanding of the factors behind online polarization, in the present work we collect and analyze Web browsing histories of tens of thousands of users alongside careful measurements of the time spent browsing various news sources.
We show that online news consumption follows a polarized pattern, where users' visits to news sources aligned with their own political leaning are substantially longer than their visits to other news sources.
Next, we show that such preferences hold at the individual as well as the population level, as evidenced by the emergence of clear partisan communities of news domains from aggregated browsing patterns.
Finally, we tackle the important question of the role of user choices in polarization. Are users simply following the links proffered by their Web environment, or do they exacerbate partisan polarization by intentionally pursuing like-minded news sources?
To answer this question, we compare browsing patterns with the underlying hyperlink structure spanned by the considered news domains, finding strong evidence of polarization in partisan browsing habits beyond that which can be explained by the hyperlink structure of the Web.
\end{abstract}

\section{Introduction}

Many people---according to the Pew Research Center, one in three Americans in 2018~\cite{geiger2019key}---prefer to receive their news online.
Market data on digital publishers further indicates that search engines and social media are dominant sources of referrals to online news content, totaling over 75\% of referral traffic~\cite{parsley}.  
Taken together, these numbers suggest that, for many typical news consumers,
the news they see is to a large part determined
by the online platforms they frequent.

At the same time, recent research \cite{rogowski2016ideology,webster2017ideological,iyengar2019origins} has highlighted polarization of political beliefs among the general population, with specific attention to the relationship between the agendas of political parties and the partisan appeal of online news organizations.
As online news consumption increases and evidence of polarization among the general population accumulates, understanding the mechanisms of partisan information\hyp seeking behaviors and the properties of online news environments grows in importance.

Not only do social media constitute an important entry point for discovering news content;
endorsements of news on social media have also been identified as a factor by which people trust news sources, presumably because endorsements represent a vote in favor of the source's credibility \cite{messing2014selective}.
Search engines---another important entry point for discovering news content---are driven by
algorithms that exhibit a certain degree of political polarization, and additionally, the input queries of individual users themselves are also polarized in terms of their partisan alignment \cite{kulshrestha2017quantifying}.  
Polarization, thus, is the product of both individual preferences and properties of the content platforms used to seek out exposure to news.

In this paper, we examine the information\hyp seeking behaviors of individual Web users in order to contribute to understanding the relationship between mass polarization and the consumption of news content on the Web.
To address this general research question, we analyzed a novel, large-scale dataset consisting of browsing logs collected via an observational field study design.  

We found evidence that supports prior research \cite{flaxman2016filter,bakshy2015exposure} indicating the existence of partisan preferences towards online news exposure, whether measured as the visit share in browsing histories or as the amount of time spent actively engaging with partisan news sources.  
Indeed, our analysis of time spent as opposed to browsing history provides more unequivocal evidence of partisan selectivity in news sources, strongly endorsing the perspective that online news consumers pursue like-minded partisan sources for political information.
We find that such polarized preferences also hold when aggregating behavior across users.

We also explore the extent to which personal preferences for news sources that cater to like-minded partisan audiences dictate polarized browsing behaviors, as opposed to the homophilic network structure of news sources on the Web.  
Since online news sources are more likely to contain links and references to like-minded partisan content, we considered the possibility that polarized browsing patterns could mostly be attributed to the nature of content options presented to a typical online news consumer.
Indeed, one of our main contributions consists of our efforts to isolate the extent to which individual partisan preferences for political information contribute to polarization patterns in news exposure, by separating the effects of user preferences from those of the biased hyperlink network of online news content.  
To do so, we compared polarization patterns in two sources of data:
a co-browsing graph consisting of the direct observation of political information\hyp seeking patterns of users,
and a hyperlink graph obtained from a crawl of said news content during the same study period.
Our analysis provides strong evidence that users specifically prefer browsing paths that lead towards polarized content, indicating that user choices play an important role in polarization.

In summary, we make the following contributions:
\begin{enumerate}
\item We collected browsing histories of tens of thousands of users in order to better understand polarization in online news consumption (Section~\ref{sec:dataset}).
\item We provide new evidence that considering the amount of time spent on news sources typically preferred by partisans reveals a different extent of polarization than observed in the prior literature, which has mostly been based on whether a user visited a site or not (Section ~\ref{subsec:coclustering} and \ref{subsec:other_side}).
\item We find that such polarization is prevalent at both the individual and the population level. We show that aggregated co-browsing graphs created from co-visited domains have a clear community structure consisting of left-, center-, and right\hyp leaning domains (Section~\ref{subsec:community}).
\item We address the question of whether polarization of news consumption can be attributed to the biased link structure of online news networks alone, or whether users' explicit content choices contribute as well. We perform a novel comparison of two distinct data sources (co-browsing choices of users vs.\ the link structure of the same content) and conclude that link structure alone cannot account for the high levels of selective exposure exhibited by online news consumers (Section~\ref{sec:user_vs_structure}).
\end{enumerate}

\section{Related work}

\xhdr{Selective exposure in online news browsing}
A popular explanation given for partisan polarization in news browsing is \textit{selective exposure}~\cite{klapper1960effects,stroud2010polarization,sears1967selective}, the observation that individuals prefer to expose themselves to information that reinforces their existing attitudes and interests. 
As a result, many have suggested that partisan preferences predict the ideological composition of news content that people will select for consumption \cite{sunstein2009republic,pariser2011filter}.  

But the impact of partisanship on selective exposure remains an open research question, as studies are mixed as to the extent to which the general population exhibits polarized browsing habits.
Much of the general population appears to consume a centrist media diet, with a long tail of more extreme partisans who exclusively visit partisan-leaning websites~\cite{nelson2017myth}.
\citet{flaxman2016filter} examined the Web-browsing histories of 50,000 US-based users of the Bing toolbar and highlighted that a majority had the highest density of visits on a few mainstream news outlets.
\citet{guess2016media} used data from the online Web\hyp tracking service YouGov and concluded that there was a small group of highly partisan and active users who drove a disproportionate amount of traffic to ideologically slanted websites. 

Various laboratory-based studies have illustrated how individual\hyp level selective exposure could explain large polarization effects in news exposure.
\citet{iyengar2009red} experimentally manipulated source labels of news sites to show preferences in sources rather than content, and showed that study participants were consistent in their selection of like-minded sources for both political and non-political information.
\citet{garrett2009echo} found selective preference among users from partisan websites, but also observed that a preference for like-minded content did not imply an aversion to challenging information; while people spent time on content that reinforced their beliefs, they also spent time on content with which they disagreed.

\citet{gentzkow2011ideological} measured polarization as the degree of ideological segregation in online and offline news exposure, as well as face-to-face interactions, and determined that the Web was no more polarized than other means of consuming political information. 
But in a field study similar to our panel design, \citet{peterson2018echo} scrutinized online browsing histories and concluded that polarized browsing habits appeared to be 2--3 times larger than previous studies had estimated.
Thus, there appears to be a lack of consensus between field and laboratory studies as to whether the Web enables greater selective exposure to political information in news content.

\xhdr{Polarization in social media and search engines} 
As news consumption moves to social media~\cite{barthel2015evolving}, researchers have started to investigate how sharing habits of news content on social media platforms are polarized along partisan lines. 
Numerous researchers have focused on polarization effects in individual interactions on social media platforms~\cite{schmidt2017anatomy,narayanan2018polarization,garimella2018political}.
For example, \citet{raghavan2018mapping} examined the structure of online interactions in order to determine how different media sources are ``invoked'' in replies on social media posts, concluding that polarization exists within the context of the way people use social media platforms outside of news exposure.

Existing research has, however, not reached a consensus on whether the use of social media networks and search engines increases the likelihood of polarized browsing habits \cite{nelson2017myth,flaxman2016filter,nikolov2019quantifying}.
For example, \citet{barbera2015tweeting} showed that polarization in news consumption on social media had been overestimated, while studies using survey data~\cite{barthel2015evolving} showed that people self-report polarized browsing habits.
\citet{bakshy2015exposure} compared the role of individual choices and algorithmic personalization on news consumption on Facebook and concluded that personal preferences, rather than content recommendations on social feeds,  drove content choices.

\xhdr{Measuring political leaning of sources and users}
Social media user behavior has also been used to identify polarity (or leaning) scores of news domains and hence to quantify polarization. \citet{bakshy2015exposure} created a list of 500 news domains and assigned a leaning score to a news domain based on the fraction of self-identified conservative users on Facebook who had shared the domain. A similar, yet much larger, list was created by \citet{robertson2018auditing}, who used a panel of Twitter users of self\hyp identified political leaning in order to compute leaning scores for sources.
Following a similar methodology, \citet{kulshrestha2017quantifying} used lists of Twitter users who were tagged as Democrats or Republicans and quantified the leaning of users based on whom they followed from these lists. Then, based on the following of various news organizations on Twitter, they identified the leaning of news domains.
\citet{ribeiro2018media} made use of Facebook's advertising platform data to compute the leaning of a news source by the extent to which liberals or conservatives were over- or under-represented among its audience.
Finally, \citet{lahoti2018joint} made use of the relationship between user preferences and their news consumption in order to jointly estimate both user leaning and news domain leaning on Twitter.

\section{Data and definitions}
\label{sec:dataset}

We randomly recruited participants from the active Firefox user population on 5~April 2018 for enrollment in an extension\hyp based study.
The sampled participants were presented with a prompt asking whether the user was interested in participating in a research study involving close scrutiny of their browsing behavior, including their browsing histories. 
The prompt and process of informed consent was approved by
Mozilla's
legal and ethical processes.
When a user consented, a browser extension was installed, which created a record logging the user's active dwell time on content organized by URL for each day of browsing activity.  
Dwell time is implemented via the \texttt{nsIdleService} API of the browser, which checks an operating system--level idle detection API every five seconds.  
In short, this method creates a log of activity per URL that is persistent across multiple tabs and windows. This is novel compared to many other methods of measuring attention to online content, which typically measure duration as a result of analytics or tracker information visible from server logs or simply browsing history without duration measurements.  If a participant opened a link but spent very little time engaging with the content of the page, browsing history--based measurements would weigh such a visit equally to a high-engagement visit.  

In contrast to visit counts, the dwell time measurement consists of three distinct states that a site visit can be in: active, idle, or not in session.  In our data collection, a visit to a website, either through a user opening a new window or a new tab, triggers an event for that URL to be logged. 
After five seconds of inactivity (measured as a lack of keyboard or mouse interaction with the browser), the site visit is considered to have entered into an idle state.  If the user resumes interaction with the content within 30 minutes, we resume our logging for the same URL.  If the user moves the focus away from the URL (e.g., by opening a new window or transferring focus to a new tab) or after 30 minutes of idle time have elapsed, the session is ended.

\begin{figure}[t]
\centering
\begin{minipage}{.9\linewidth}
\centering
\subfloat[Number of events]{\label{fig:activity}\includegraphics[width=\textwidth, height=0.6\textwidth, clip=true, trim=0 60 0 60]{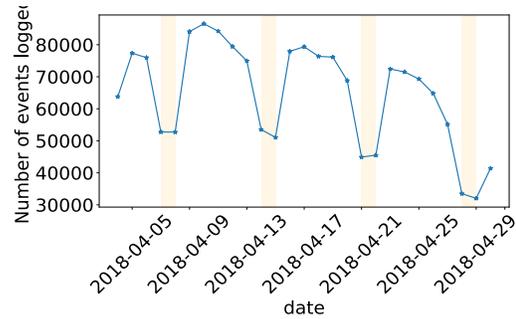}}
\end{minipage}%
\par\medskip
\begin{minipage}{.9\linewidth}
\centering
\subfloat[Average dwell time]{\label{fig:dwelltime}\includegraphics[width=\textwidth, height=0.6\textwidth, clip=true, trim=0 60 0 60]{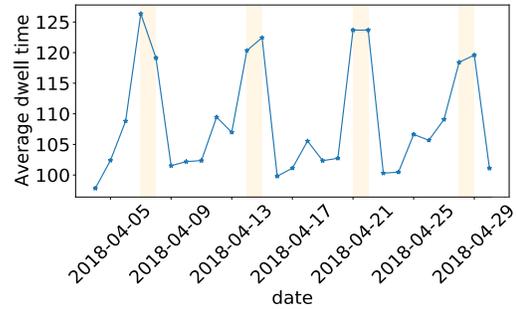}}
\end{minipage}%
\caption{
(a) Daily number of events logged during the three-week study period.
(b) Average dwell time per page visit (in seconds).
The shaded bands indicate weekends.
}
\label{fig:activity_dwelltime}
\vspace{-\baselineskip}
\end{figure}

\xhdr{Activity statistics of study participants}
Our dataset consists of browsing logs from a panel of 24,036 
unique participants, based in the US, for a period of three weeks starting on 5~April 2018. The dataset consists of 241 million distinct page visits with a median of 6,074 page visits per individual.
Figure~\ref{fig:activity} shows the activity of all users.
We observe that activity (number of events logged) decreases over the course of the study period, and that there is a noticeable drop during weekends. 
On the contrary, Figure~\ref{fig:dwelltime} shows that the average dwell time per visit is larger on weekends than during the week, and that, even as the number of events logged decreases over the course of the study period (Figure~\ref{fig:activity}), this is not the case for the average dwell time. 
This suggests that, despite participant churn, the general pattern of interactions remains consistent.
For robustness, we repeated all analyses on the subset of users who remained active throughout the observation period, finding no qualitative disagreement from the results obtained on the set of all users.

\xhdr{Bias and limitations of dataset}
Since the data was constructed from the browsing activity of a panel of opt-in Firefox users, we acknowledge that the data is likely to be biased.
As we do not have demographic covariates for members of the study panel, we are not able to compare the representativity of this sample against more conventional measures of the general population, as \citet{guess2016media} and \citet{peterson2018echo} were able to do.
Since it is possible that the preference to choose Firefox as a primary browser could be associated with one's latent news preferences (in that political predispositions could predict both the likelihood to use Firefox and exposure to news content), we considered the possibility that the participants in this study might deviate from the general population in their overall news preferences.  
We compared the top 100 most visited news sites in our Firefox logs with the top 100 most popular (with respect to unique monthly US visitors) news sites according Alexa.com, a prominent website ranking service.
Spearman's rank correlation between the two lists was 0.85, suggesting that our study participants were generally exposed to similar popular sources of online content as the general population. 

\xhdr{Partisan leaning computation}
All our analysis is at the domain level.
We obtained partisan leaning scores for all domains from \citet{robertson2018auditing}, where ``leaning'' is defined as the estimate of a news outlet's political ideological alignment with either a conservative or a liberal audience.
The leaning score of a domain assumes a value between 0 and~1;
it is defined as the average self-reported ideology of users who have shared pages from that domain on Twitter,
where 0 corresponds to maximally liberal, and 1 to maximally conservative.
The original list provided by \citet{robertson2018auditing} has leaning scores for over 19,000 domains, including domains such as facebook.com, google.com, and instagram.com.
However, since we are specifically interested in understanding news consumption behavior, we only used a subset of news domains, obtained from the popular journalism watch-dog website \url{mediabiasfactcheck.com}.
The subset contained 2,873 news domains that were still online in September 2020, and further discarding domains without a leaning score in Robertson et al.'s list resulted in a set of 1,295 news domains.

\begin{figure}[t]
\centering
\begin{minipage}{.9\linewidth}
\centering
\subfloat[Histogram of domain leaning scores]{\label{}\includegraphics[width=\textwidth]{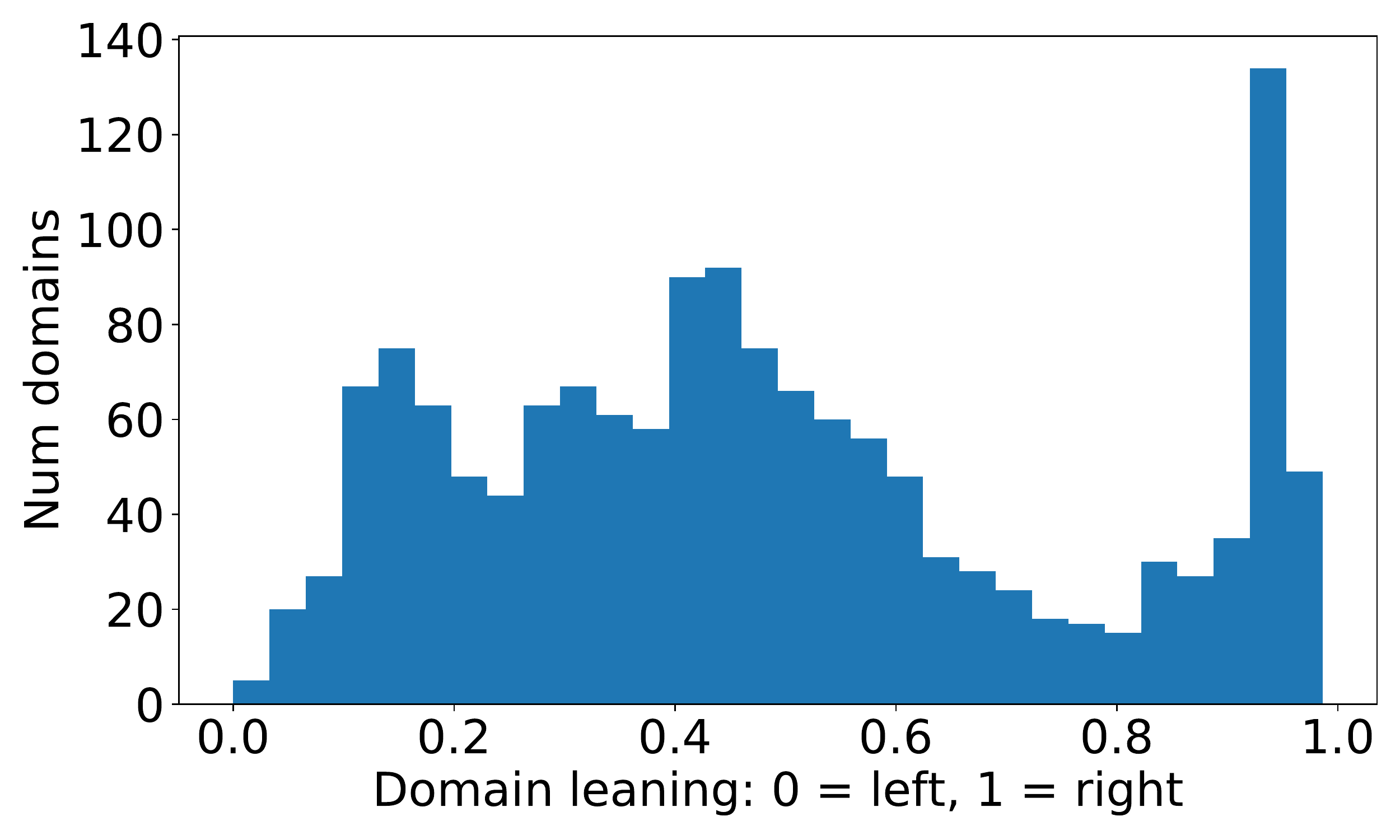}}
\end{minipage}%
\par\medskip
\begin{minipage}{.9\linewidth}
\centering
\subfloat[Histogram of user leaning scores]{\label{}\includegraphics[width=\textwidth]{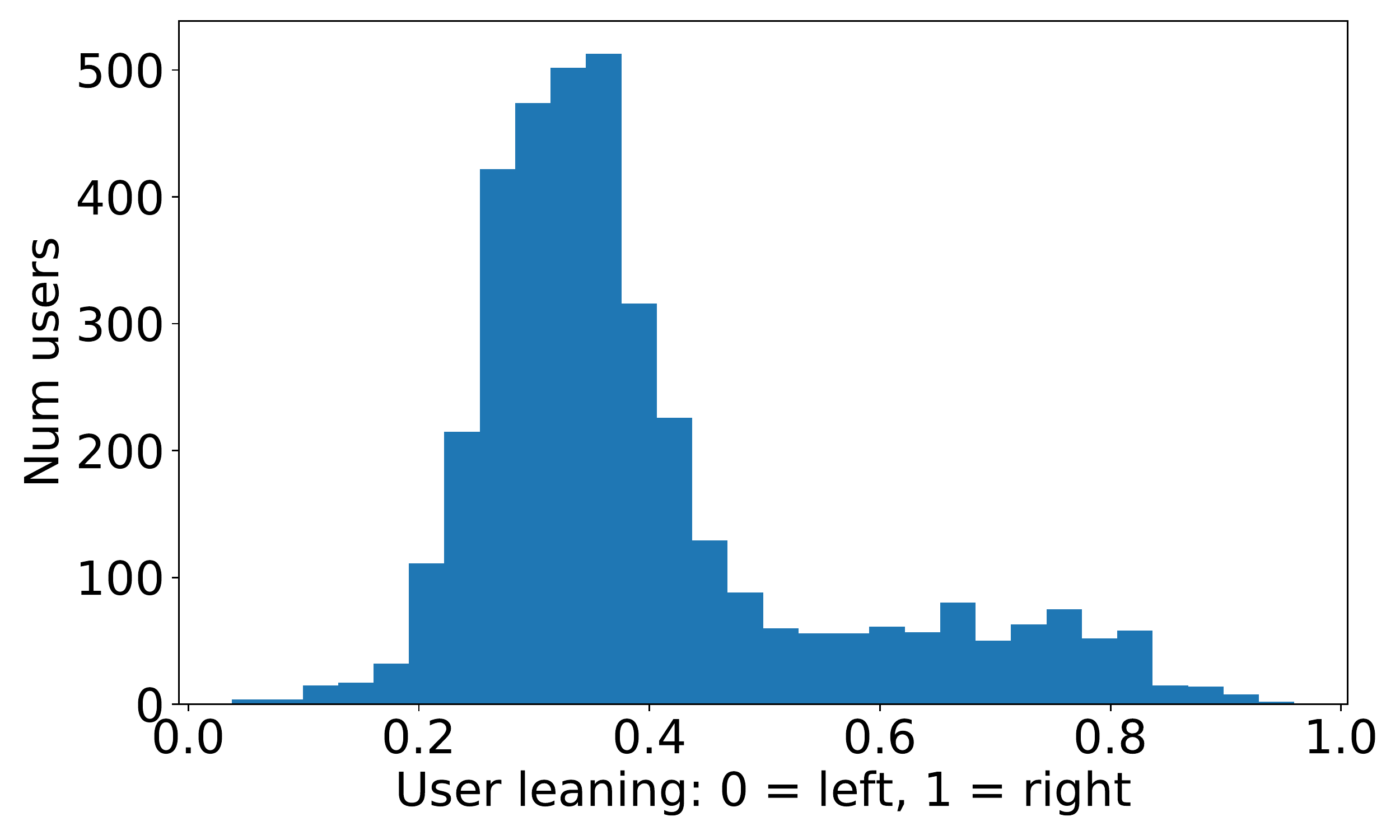}}
\end{minipage}%
\caption{
Histogram of leaning scores for
(a)~domains and
(b)~users.
}
\label{fig:leaning_domains}
\vspace{-\baselineskip}
\end{figure}

One should bear in mind that defining the leaning of a domain based on the fraction of self\hyp identified users with a specific political leaning could be biased due to self\hyp selection.
Although we acknowledge this bias, we argue that any assignment of a leaning score to a domain (say, by a journalist or by a professional organization) will be subject to similar personal biases.
The leaning scores we use have been shown to be highly correlated with the scores from \citet{bakshy2015exposure}, who used a similar methodology on Facebook, thus providing some evidence that the scores are robust. We use the list of \citet{robertson2018auditing} because it is larger and more up-to-date.
The 1,295 studied domains include a mix of news sources (e.g., nytimes.com, washingtonpost.com), cable TV sites (e.g., msnbc.com, foxnews.com), etc.
The distribution of the leaning scores of domains is shown in Figure~\ref{fig:leaning_domains}a.




In this study, we only consider participants in our analysis who
in total made at least 50 visits to pages from the 1,295 considered news domains.
This threshold restricted the dataset to 6,575 participants.
We compute the political leaning of individual participants as the weighted average leaning score of the domains they visited, where each domain was weighted by the number of times it had been visited by the respective user.
The distribution of participants' leaning is shown in Figure~\ref{fig:leaning_domains}b.
We can see that, on average, the studied participants are leaning liberal, as estimated through their online news browsing history.
Note that, in the rest of the paper, we shall use the terms leaning\slash polarity\slash ideology, left-leaning\slash liberal, and right-leaning\slash conservative interchangeably.



\xhdr{Browsing graph}
Important parts of our analysis are based on the bipartite (user-by-domain) \textit{browsing graph}
$G$ = $(U,D,E)$,
where $U = \{u_1, \dots, u_n \}$ are the $n$ users,
$D = \{d_1, \dots, d_m\}$ are the $m$ news domains,
and $E$ is the set of edges $\{\{u_i, d_j\} : u_i \in U, d_j \in D\}$ between users and domains.
Each edge $\{u_i, d_j\}$ is associated with a weight $w_{ij}$ that captures the average dwell time spent by user $u_i$ on pages from domain $d_j$ (e.g., if a user on average spends 31 seconds on pages of \url{nytimes.com}, the corresponding weight is 31).
The bipartite graph $G$ may also be thought of as a matrix whose rows are users, whose columns are news domains, and whose entries indicate the average dwell time of a user on a domain.

\xhdr{Co-browsing graph}
Using the bipartite browsing graph $G$, we performed a one-mode projection onto the news domains $D$ by counting all paths of length 2 between two domains.
We call the graph resulting from the projection the \textit{co-browsing graph.}
Edges between two domains $d_1$ and $d_2$ in the co-browsing graph are weighted in order to represent the number of users who visited both $d_1$ and $d_2$.
To count co-visits of domains,
we aggregated URLs at the domain level,
considered only those domains for a given user with which they had actively engaged for at least 60 seconds in total (measured via the sum of the user's dwell times on the domain),
and counted multiple visits by the same user to the same domain only once.
Summary statistics of the co-browsing graph are listed in Table~\ref{tab:cobrowse}.

\xhdr{Hyperlink graph}
In order to investigate users' browsing patterns in the light of how online news websites link to one another, we constructed the \textit{hyperlink graph} of the network of news domains.
The Common Crawl project\footnote{\url{http://commoncrawl.org/}} regularly crawls the Web and releases datasets of the discovered webpages on a regular schedule, which are generally considered the most extensive publicly available sources regarding the structure of the Web~\cite{meusel2014graph}.
Common Crawl also periodically releases versions of the hyperlink graph of the Web aggregated at the domain level, specifying for each domain to which other domains it contains hyperlinks.\footnote{e.g., \url{http://commoncrawl.org/2019/02/host-and-domain-level-web-graphs-nov-dec-2018-jan-2019/}}
Our analysis works at the domain level and is restricted to the subgraph defined by the set of 1,295 news domains $D$ also used in constructing the browsing and co-browsing graphs.
In order to enable a meaningful comparison with the co-browsing graph, we ignore the directionality of edges and consider the hyperlink graph as an undirected graph.
The hyperlink graph represents static connections between domains, and results should be interpreted in this framework.
Summary statistics of the hyperlink graph are listed in Table~\ref{tab:cobrowse}.
In particular, we observe that the hyperlink graph is denser than the co-browsing graph.

\begin{table}[]
\smaller
\caption{Statistics of co-browsing and hyperlink graphs.}
\label{tab:cobrowse}
\begin{tabular}{l|l|l|l|l}
\hline
	  & \textbf{Nodes} & \textbf{Edges}  & \begin{tabular}[c]{@{}l@{}}\textbf{Avg.}\\ \textbf{degree}\end{tabular} & \textbf{Top PageRank}                                                                         \\
\hline
Co-browsing & 1,295   & 176,945 & 273                                                  & \begin{tabular}[c]{@{}l@{}}nytimes.com,\\ washingtonpost.com,\\ cnn.com,\\ theguardian.com,\\ npr.com, \end{tabular}           \\
\hline
Hyperlink & 1,295  & 323,036 & 498                                                   & \begin{tabular}[c]{@{}l@{}}washingtonpost.com, \\ huffingtonpost.com, \\ nytimes.com,\\ businessinsider.com,\\ theatlantic.com 
\end{tabular} \\
\hline
\end{tabular}
\vspace{-\baselineskip}
\end{table}

The browsing, co-browsing, and hyperlink graphs capture different aspects of how users browsing the Web move from domain to domain.
The bipartite browsing graph reflects choices made by users; it aggregates each user's entire browsing history.
The co-browsing graph further aggregates across all users, thus giving a macro picture of the browsing patterns of online news.
In contrast, the hyperlink graph reflects the choices made by content creators to link between pages.
In this sense, the hyperlink graph also represents the set of possible direct links that users \textit{could} have taken while browsing.

\section{Polarization patterns in news browsing}
\label{sec:polarization}


In this section, we build on top of the above-described datasets to characterize polarization in news browsing. We start by investigating dwell times spent on various domains as a way of showing the existence of polarization. Next, we analyze communities of domains in the co-browsing graph in order to show that certain groups of news domains tend to be co-visited by the same user.

\subsection{Polarization in dwell times}
\label{subsec:coclustering}
First, we assess the extent of polarized browsing habits among participants by examining whether there is any clear structure in the bipartite browsing graph through the use of spectral graph co-clustering techniques~\cite{dhillon2001co,role2018coclust}.
Unlike graph clustering techniques, which can only cluster one dimension (users or domains) independently, co-clustering techniques find the dependencies between participants and domains.
Recall that edge weights of the bipartite browsing graph capture each user's average dwell time for each domain (the average amount of time they spent actively engaging with pages from the respective domain), 
such that, by co-clustering the bipartite browsing graph, we obtain clusters of similar domains visited by similar users.
To avoid individual activity biases, a given user's domain-level dwell times were $z$-score-standardized across all domains they had visited.
Given this setup, polarized browsing habits would appear as tight clusters of domains with average domain leaning scores that significantly differ across clusters.

\begin{figure}[t]
\begin{minipage}{.9\linewidth}
\centering
\subfloat[Weighted browsing graph]{\label{}
	\includegraphics[width=\textwidth]{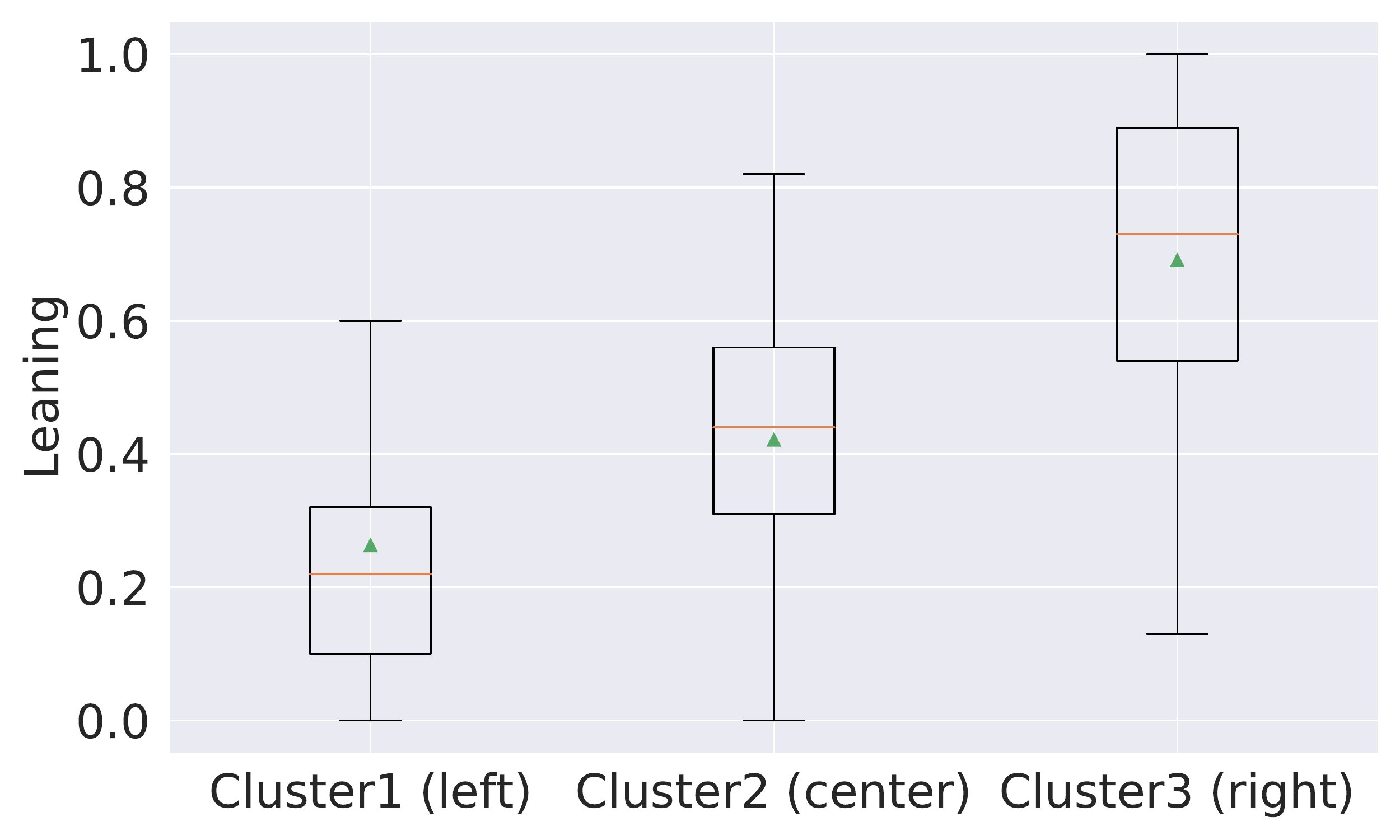}}
\end{minipage}%
\par\medskip
\begin{minipage}{.9\linewidth}
\centering
\subfloat[Unweighted browsing graph]{\label{}
	\includegraphics[width=\textwidth]{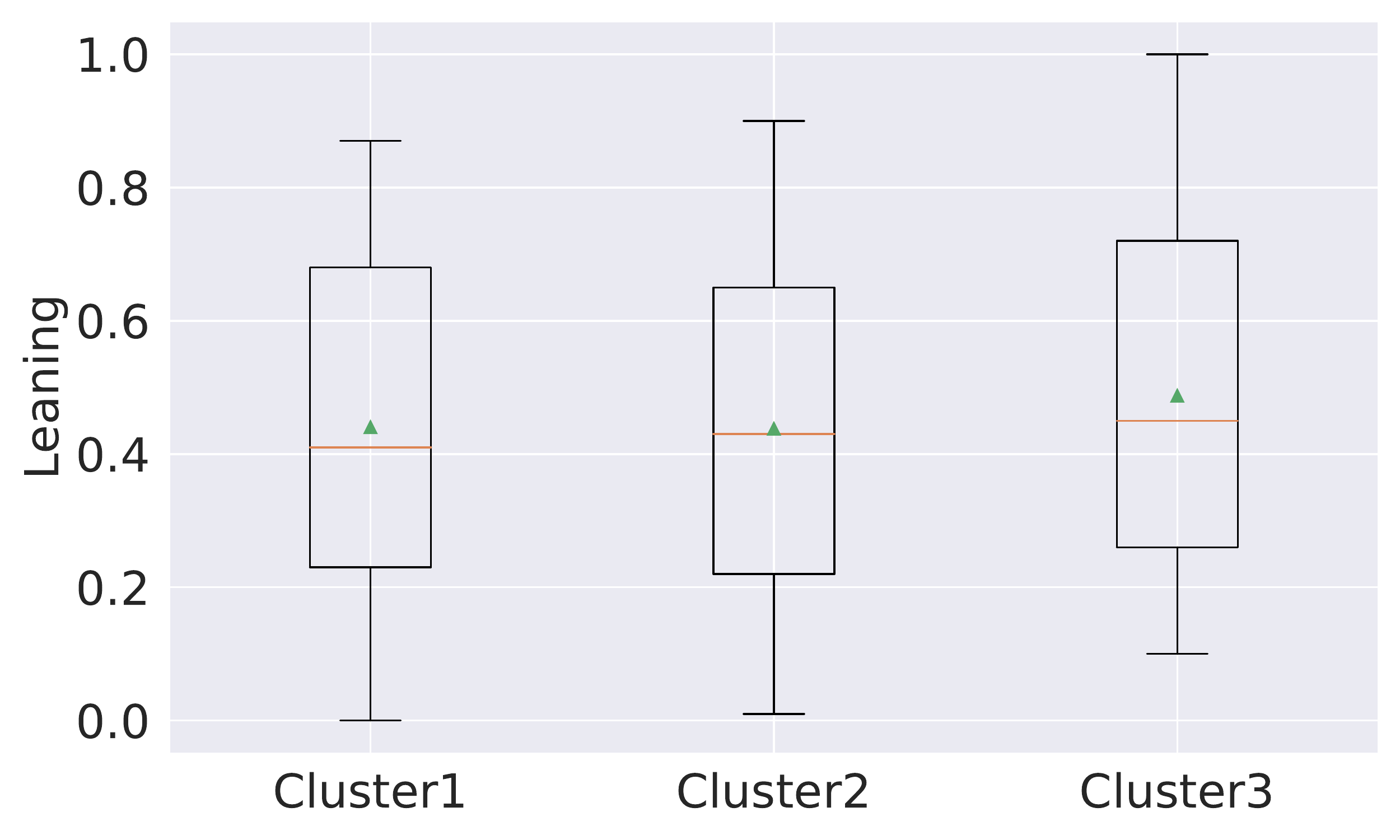}}
\end{minipage}%
    \caption{
    Boxplots of leaning scores per domain cluster, obtained by co-clustering
    (a)~the weighted bipartite (user-by-domain) browsing graph,
    (b)~an unweighted version of the browsing graph.
    Green dots are means, red lines are medians, box boundaries are quartiles.
    }
    \label{fig:coclustering}
\vspace{-\baselineskip}
\end{figure}

Running the co-clustering procedure for a number of co-clusters ranging from two to five, we found that when using three co-clusters, a liberal-leaning (left), a centrist, and a conservative-leaning (right) cluster of domains emerged.
Since, moreover, using more than three co-clusters merely split one of the co-clusters into two, we decided to set the number of co-clusters to three.
For each cluster of domains and for each cluster of users, we computed the distribution of their leaning scores.
Figure~\ref{fig:coclustering}a shows a box plot of the distribution of domain cluster leanings.
Similar trends emerged in the leaning of user clusters (not shown).

Recall that the co-clustered browsing graph was based on dwell time information only; neither domain nor user leanings were used in the co-clustering process.
In this light, Figure~\ref{fig:coclustering}a, which shows that the cluster-specific distributions of leaning scores differ widely across domain clusters, implies that domains that are visited in similar ways by similar users tend to be similar with respect to their leaning---a clear indication of partisan polarization in browsing.

In order to analyze whether the clustering structure can be identified by website visits alone (without considering dwell times), we also performed a co-clustering on an unweighted version of the bipartite browsing graph.
As observed in Figure~\ref{fig:coclustering}b, the separation in the communities is not as pronounced in the unweighted as in the weighted version, suggesting that dwell time--based measurements reveal stronger polarization patterns.
Given our novel methodology to compute the dwell time of users, this is the first large-scale
study
to establish the existence of polarization in terms of time spent by users on partisan websites as opposed to browsing histories alone.





\subsection{Time spent ``on the other side''}
\label{subsec:other_side}

Next, we investigate
whether users of different leanings engage differently with domains of different leanings.
To do so, we first split the set of news domains into left-leaning (leaning below 0.4), center (leaning between 0.4 and 0.6), and right-leaning (leaning above 0.6) domains.
Then, for each user, we computed their average dwell time for each domain and $z$-score-standardized their average dwell times across domains (such that, for a fixed user, the mean across domains is~0, with a standard deviation of~1), in order to remove effects due to the fact that some users generally spend more, and some less, time engaging with visited pages.
Next, we binned users into five buckets based on their leanings and computed the mean standardized dwell time in each user bucket, separately for the three leaning-based domain groups.

\begin{figure}[t]
\centering
	\includegraphics[width=\columnwidth]{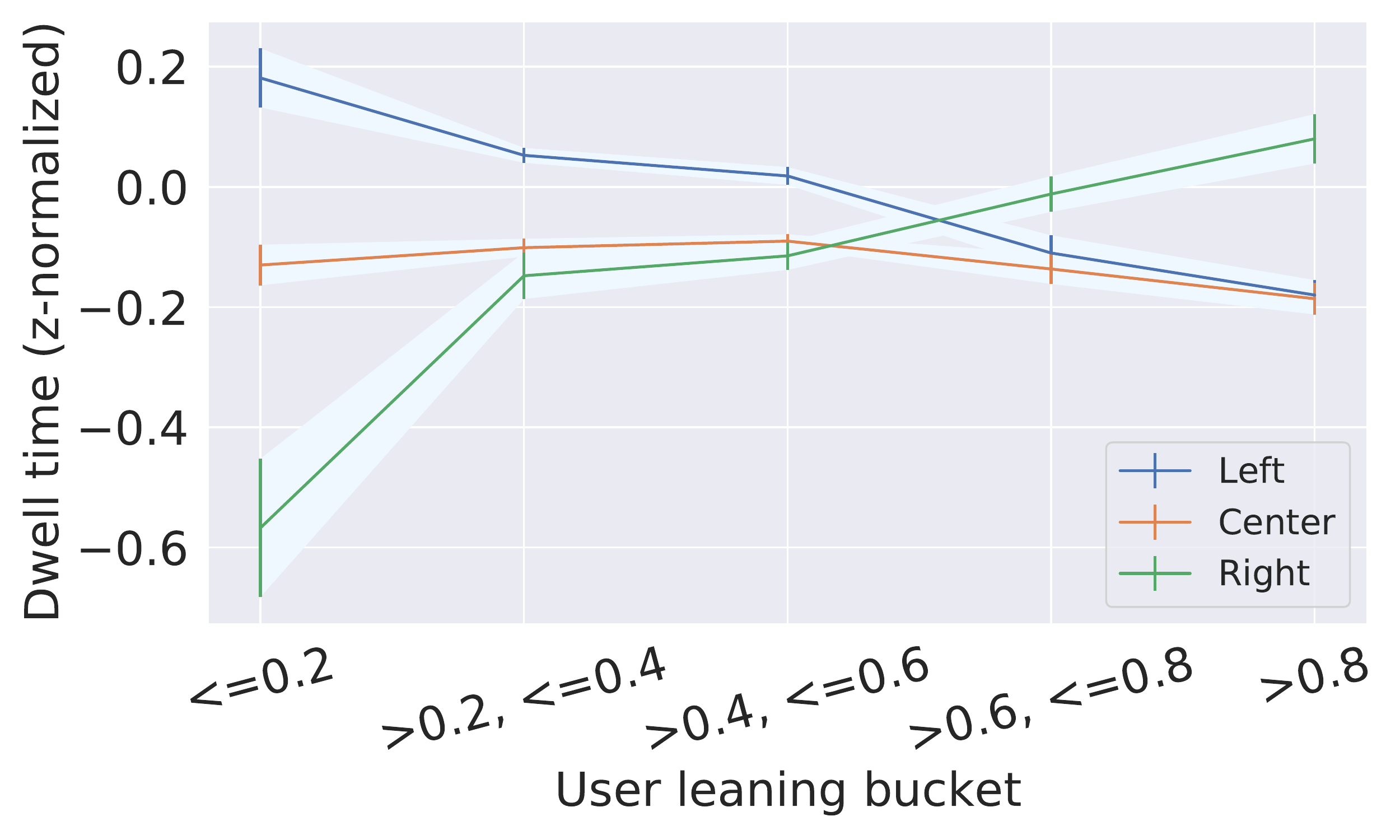}
\caption{
Average dwell time ($z$-score-standardized within users) spent by users of different leanings ($x$-axis) on domains of different leanings (three curves).
Error bars show 95\% confidence intervals.
We see that users spend significantly more time when visiting pages on domains aligned with their own leaning.
}
\label{fig:dwell_time_leaning}
\end{figure}

Figure~\ref{fig:dwell_time_leaning} shows the dwell time for left-leaning, center, and right-leaning sites for users with varying leanings.
We observe that the more extreme participants on either side of the leaning spectrum dwell significantly longer when visiting like-minded online news content.
For domains without a clear leaning (``center''), there is no such upward or downward trend.
Although \citet{garrett2009echo} observed in a lab experiment that people spend \textit{more}---rather than less---time on certain types of content from the other side, we emphasize that our population is several orders of magnitude larger than theirs and that our data was collected ``in the wild'', outside of a lab setting.

\subsection{Community structure in co-browsing graph}
\label{subsec:community}

Above, we analyzed \textit{individual} visits to news domains by participants and found patterns of polarization in the dwell time spent upon visits of domains of various ideological alignment.
%
Now, we consider browsing patterns \textit{across participants} by analyzing the community structure of the co-browsing graph.
The co-browsing graph represents users' choices in what domains are co-visited,
so if people only select a specific set of domains that appeal to their partisan ideology, we should observe communities of ideologically similar domains in the co-browsing graph.
%

\begin{figure}[t]
\centering
\begin{minipage}{.9\linewidth}
\centering
\subfloat[Co-browsing graph]{\label{}\includegraphics[width=\textwidth, height=0.6\textwidth]{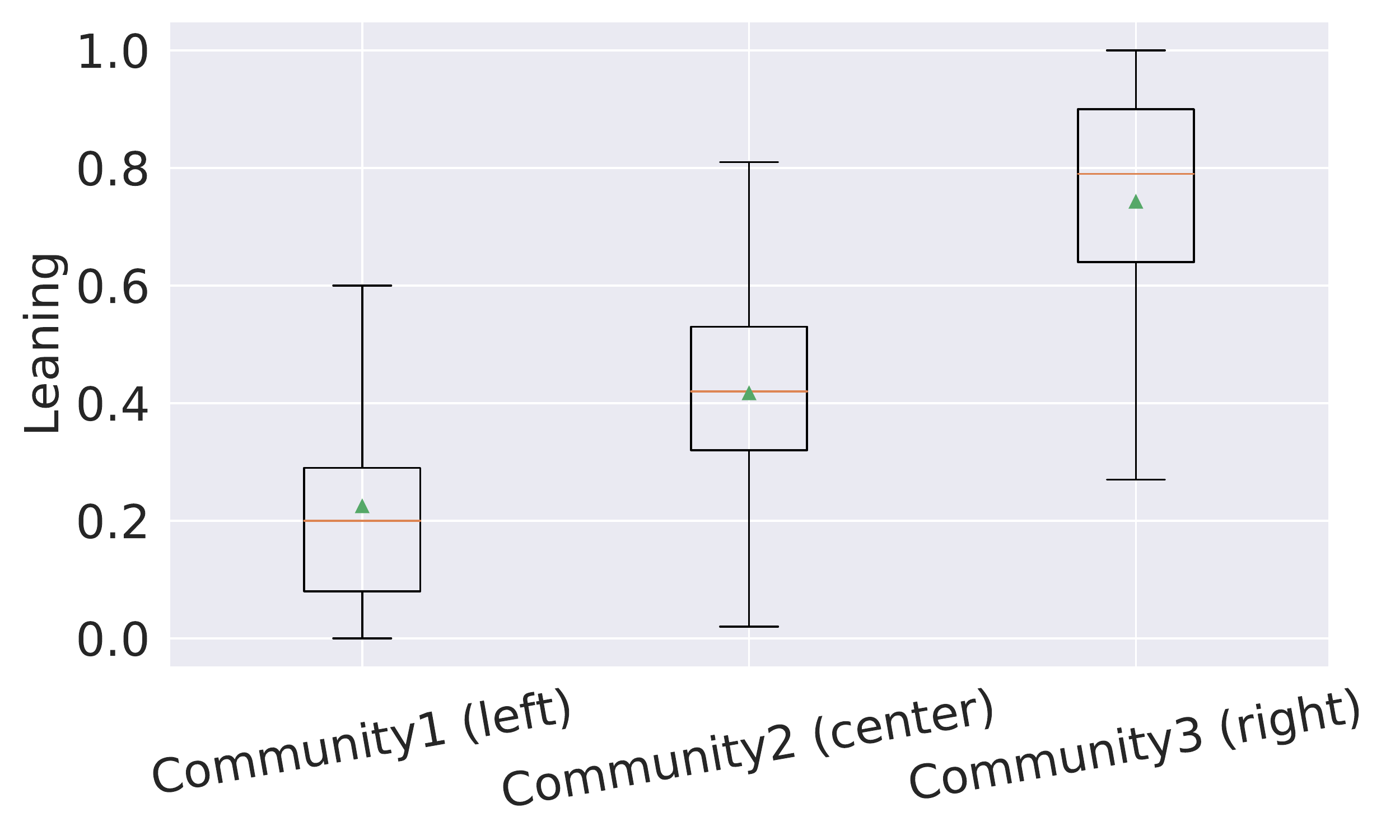}}
\end{minipage}%
\par\medskip
\begin{minipage}{.9\linewidth}
\centering
\subfloat[Hyperlink graph]{\label{}\includegraphics[width=\textwidth, height=0.6\textwidth]{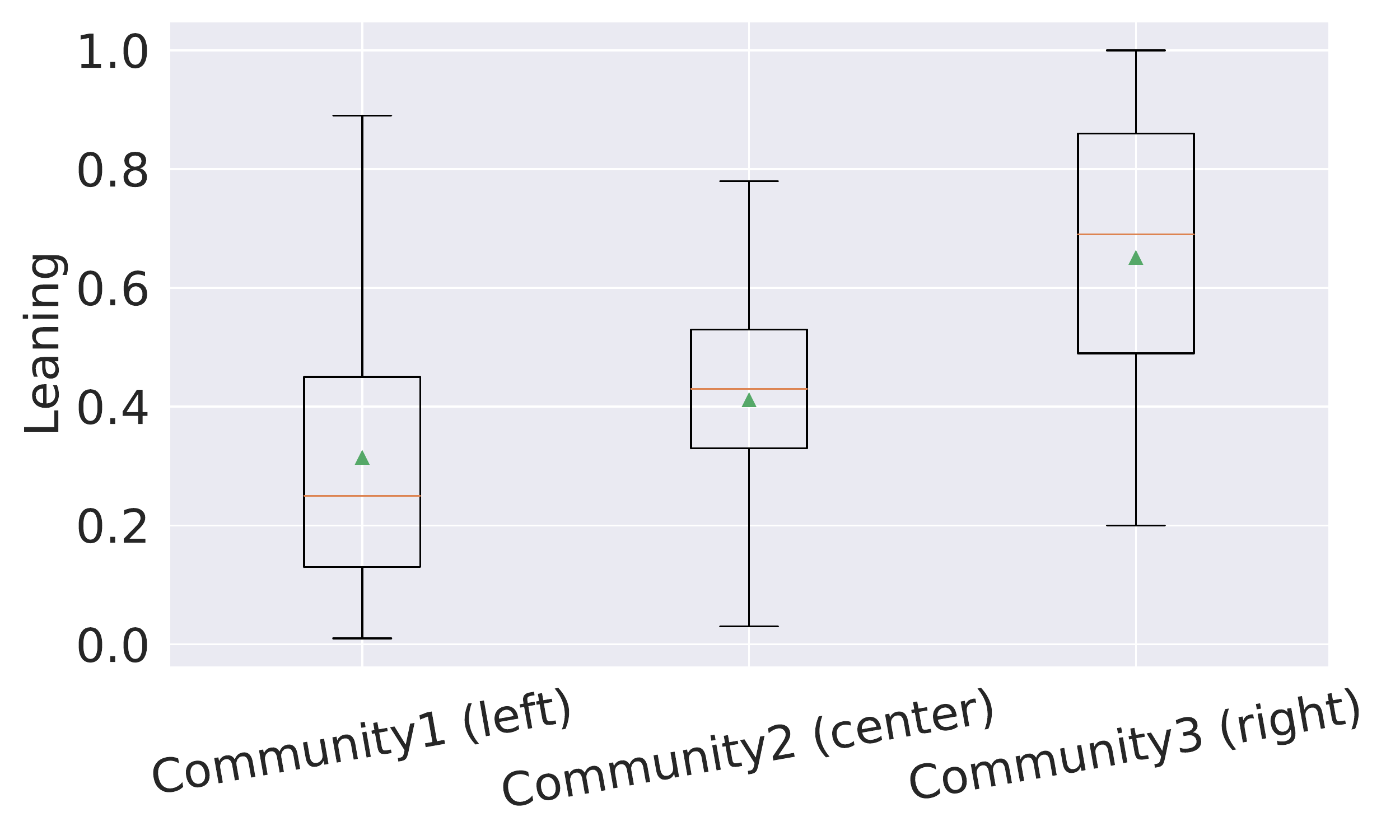}}
\end{minipage}%
\caption{
    Boxplots of leaning scores per community of domains, obtained by running community detection on
    (a)~the co-browsing graph,
    (b)~the hyperlink graph.
    Green dots are means, red lines are medians, box boundaries are quartiles.
    }
\label{fig:communities}
\end{figure}

To determine whether this is the case, we applied a community detection algorithm based on the Louvain method \cite{de2011generalized} to the co-browsing graph, producing a partition of the graph nodes into communities that maximized modularity.
Modularity measures the quality of a partition and is high if nodes are densely connected inside communities, and sparsely connected across communities.
The method, which determines the number of communities automatically, identified three communities.
Figure~\ref{fig:communities}a plots the distribution of domain leanings for each of the three communities.
Although domain leanings were not used in the community detection process, a clear stratification of the three communities along partisan lines emerges, with clusters 1, 2, and 3 containing mostly left-leaning, centrist, and right-leaning domains, respectively.
Table~\ref{tab:kcore} shows the five domains with the highest
PageRank centrality
in each of the communities. 

We also used a $k$-core decomposition~\cite{batagelj2003m} of the co-browsing graph to confirm this structure and found the same three strongly connected cores (communities), also separated along the ideological spectrum. 

\begin{table}[t]
\caption{Top five domains with respect to PageRank centrality in each community of the co-browsing graph.}
\label{tab:kcore}
\begin{tabular}{lll}
\hline
\textbf{Community 1} & \textbf{Community 2} & \textbf{Community 3} \\ 
\hline
nytimes.com                                & theguardian.com                               & foxnews.com                       \\
cnn.com                                    &  theatlantic.com
&  wsj.com                             \\
nbcnews.com                                &  thinkprogress.org                             &   breitbart.com                       \\
latimes.com                               &  motherjones.com                      
& drudgereport.com                           \\
cbsnews.com                                &        dailykos.com                         &       dailywire.com                    \\
\hline
\end{tabular}
\end{table}

In addition to the co-browsing graph, we also ran community detection on the hyperlink graph.
Here, too, three communities emerge; the results are shown in  Figure~\ref{fig:communities}b.
Although the three communities of the hyperlink graph are qualitatively similar to those of the co-browsing graph (left, center, right), the stratification is not as clear in the hyperlink graph as in the co-browsing graph (Figure~\ref{fig:communities}a):
the modularity of the co-browsing graph is twice that of the hyperlink graph (0.11 vs.\ 0.05).




\section{Selective exposure vs.\ structure of the Web}
\label{sec:user_vs_structure}

Finally, we explore whether polarization in browsing is due to selective information seeking on behalf of participants or whether it is the result of the link structure of the Web environment with which they interact.

\begin{figure}[t]
\centering
\begin{minipage}{.9\linewidth}
\centering
\subfloat[Co-browsing graph]{\label{}\includegraphics[width=\textwidth]{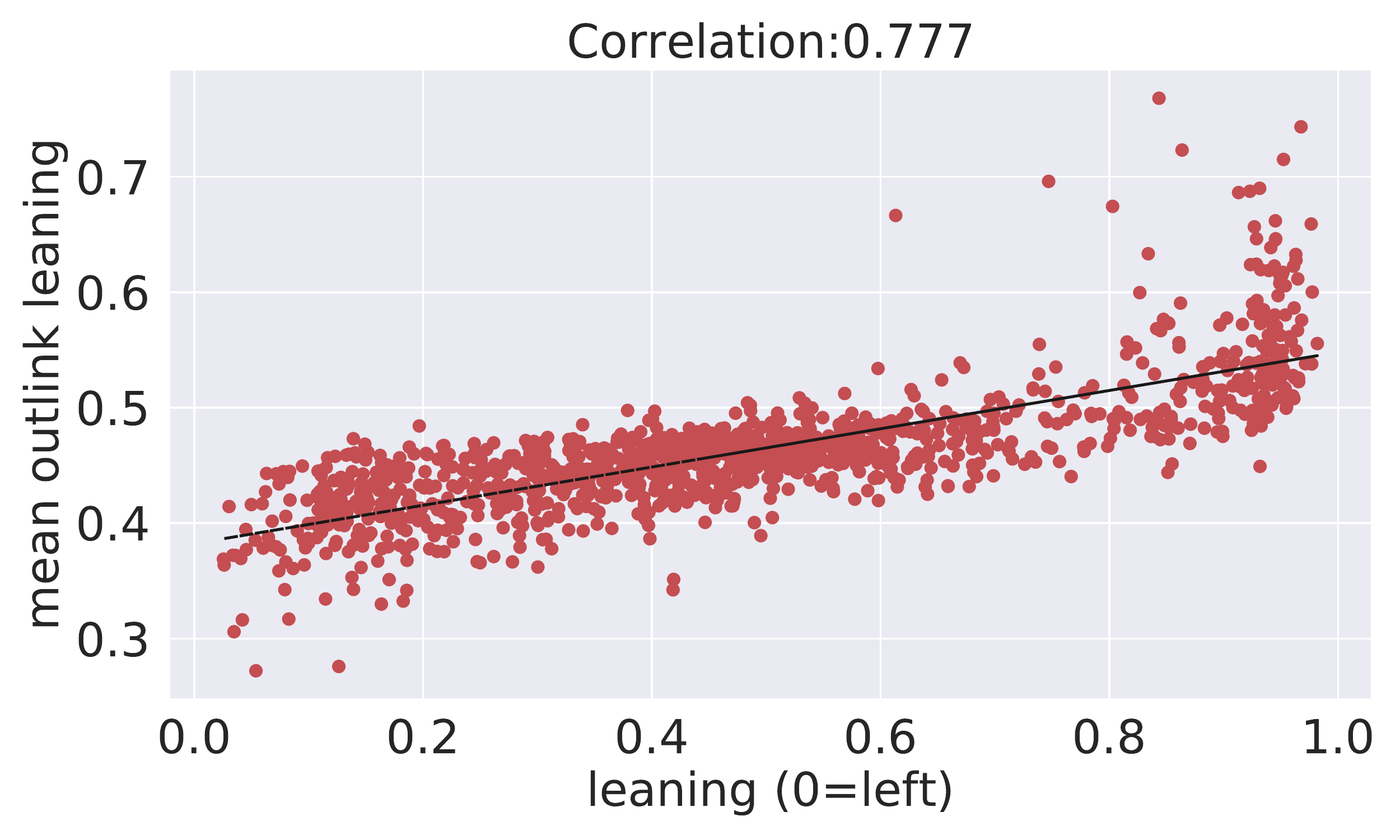}}
\end{minipage}%
\par\medskip
\begin{minipage}{.9\linewidth}
\centering
\subfloat[Hyperlink graph]{\label{}\includegraphics[width=\textwidth]{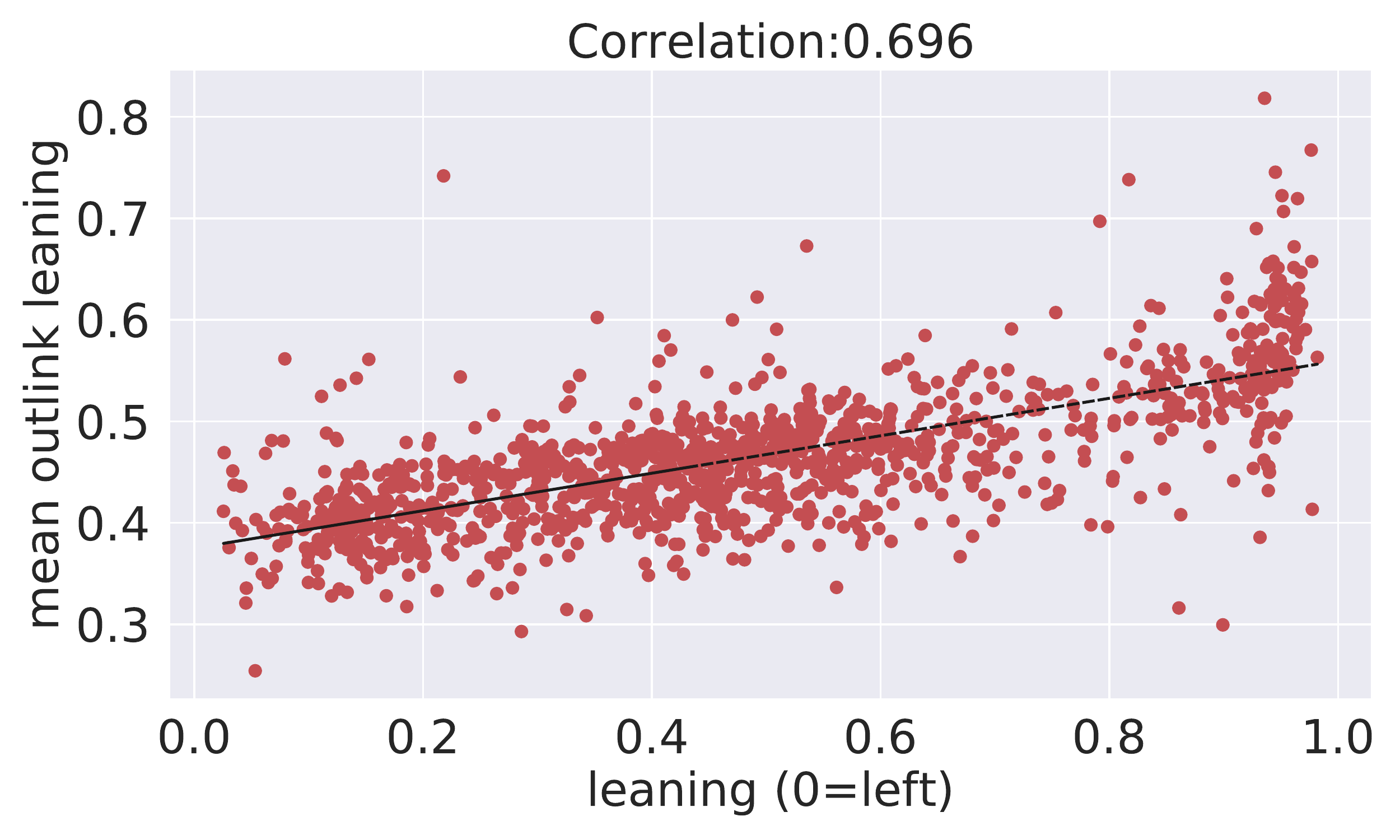}}
\end{minipage}%
\caption{
Leaning of each domain ($x$-axis) vs.\ average leaning of the domain's neighbors ($y$-axis) in
(a)~the co-browsing graph and
(b)~the hyperlink graph.
}
\label{fig:outlink}
\end{figure}

\xhdr{Homophily in hyperlink and co-browsing graphs}
To begin, we compute, for both the co-browsing graph and the hyperlink graph,
the ``neighborhood leaning'' of each node (i.e., domain) $d$, defined as the average leaning of all nodes linked to $d$ by an edge. (Recall that both the co-browsing graph as well as the hyperlink graph are treated as undirected graphs.)
Figure~\ref{fig:outlink}a, computed on the co-browsing graph, plots one point per domain $d$, showing $d$'s own leaning on the $x$-axis, and $d$'s neighborhood leaning on the $y$-axis.
We observe that, as we move from left to right, the neighborhood leaning increases, a clear sign of polarization in co-browsing patterns.
Figure~\ref{fig:outlink}b, which was computed on the hyperlink graph rather than on the co-browsing graph, exposes a similar pattern, but with a significantly lower Pearson correlation coefficient (0.696 vs.\ 0.777, $p<0.01$).
The higher correlation coefficient for the co-browsing graph indicates that neighborhoods are more homophilic in the co-browsing graph, compared to the hyperlink graph.

As a caveat, we point out that, as seen in Table~\ref{tab:cobrowse}, the average degree of nodes in the hyperlink graph is higher than that in the co-browsing graph, which might
bias the neighborhood leaning scores:
a higher average degree could \textit{a priori} lead to a more diverse set of neighbors, which could in turn lead the hyperlink graph to have less leaning homophily.

\begin{figure}[t]
\centering
\begin{minipage}{.9\linewidth}
\centering
\subfloat[Hyperlink graph vs.\ unweighted co-browsing graph]{\label{}\includegraphics[width=\textwidth, height=0.66\textwidth]{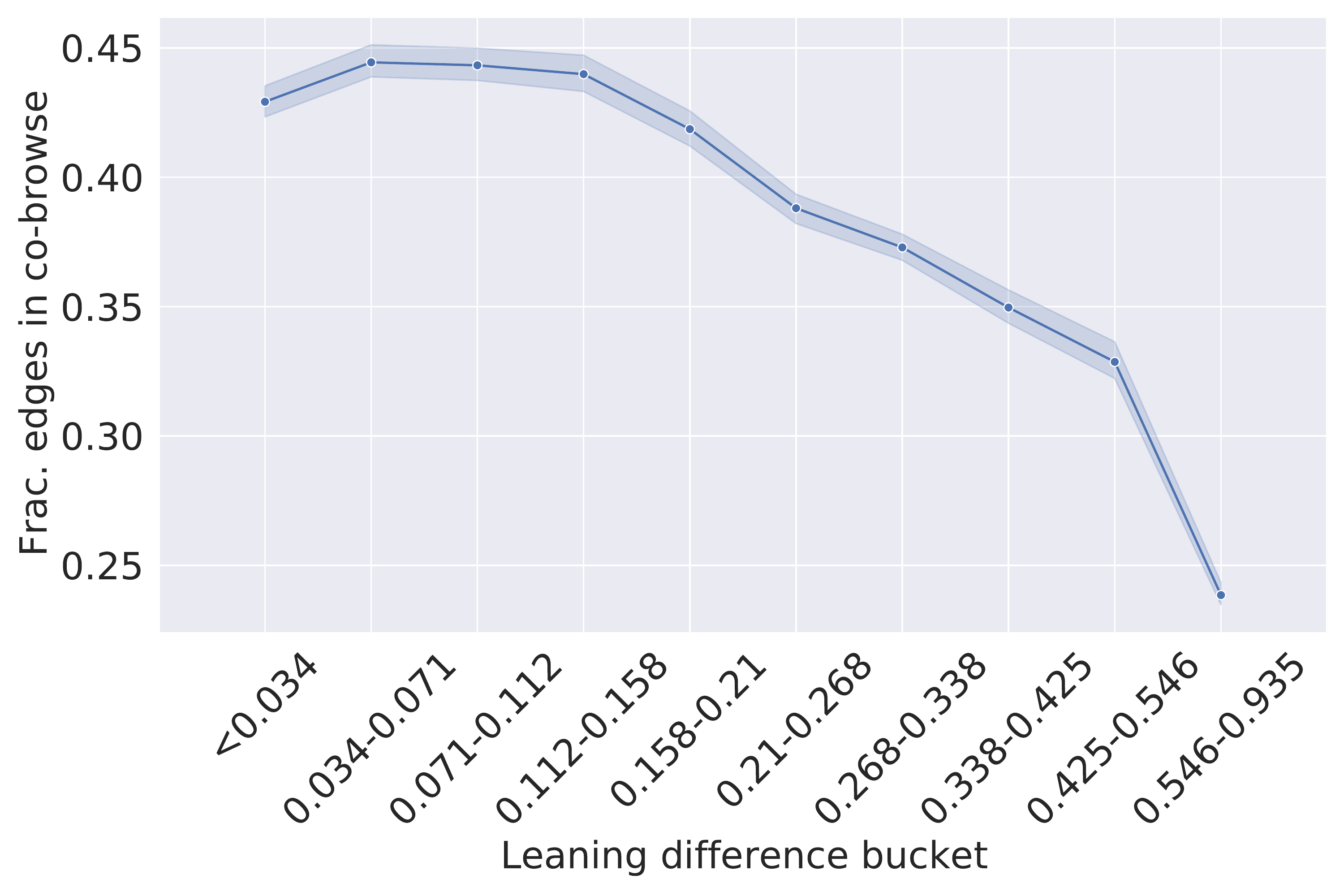}}
\end{minipage}%
\par\medskip
\begin{minipage}{.9\linewidth}
\centering
\subfloat[Hyperlink graph vs.\ weighted co-browsing graph]{\label{}\includegraphics[width=\textwidth, height=0.66\textwidth]{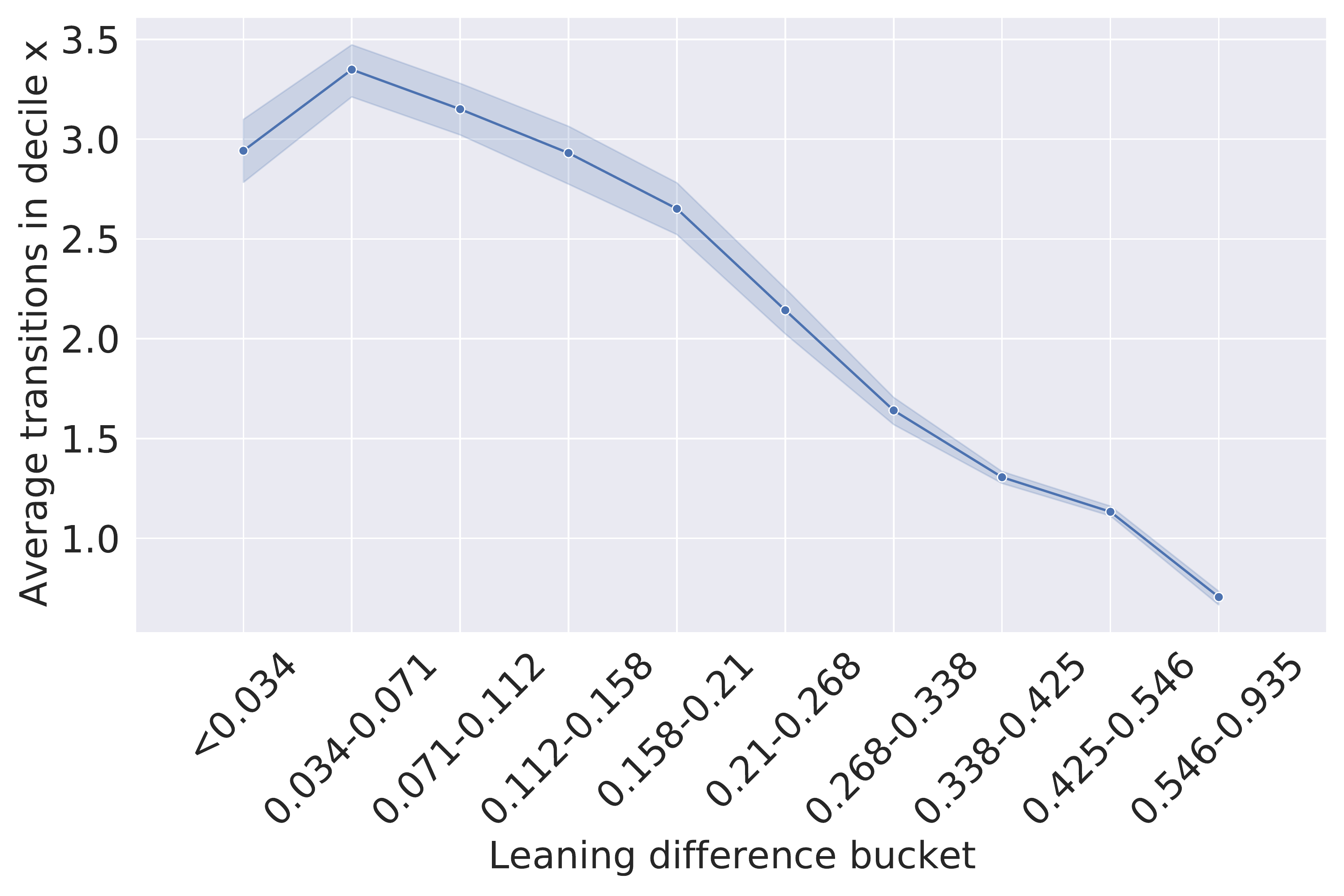}}
\end{minipage}%
\caption{
(a)~Fraction of edges from the hyperlink graph that are also present in the co-browsing graph,
when considering, for each node, only those edges that connect the node to a neighbor whose leaning difference falls into the respective decile ($x$-axis; deciles were computed per node over all its neighbors).
(b)~Analogous analysis, but with average edge weights, rather than fraction of edges, on the $y$-axis.
The plots show that hyperlinks leading to more similar (with respect to leaning) neighbors are more likely to be chosen by users.
}
\label{fig:missing_edges}
\end{figure}

\xhdr{Presence of neighbors in hyperlink versus co-browsing graphs}
Next, we compare neighborhoods of nodes in the co-browsing and hyperlink graphs.
Intuitively, an edge being present in the hyperlink graph but absent from the co-browsing graph
corresponds to participants deliberately avoiding the link when browsing.
To investigate such situations more closely, we computed,
for each node in the hyperlink graph, the absolute difference in leaning from each of its neighbors.
We then ranked the neighbors by this difference and bucketed them into deciles (only considering nodes with more than 10 neighbors, which eliminated five nodes from the analysis).
Next, for each decile, we computed the fraction of edges present in the co-browsing graph.
Our hypothesis was that, in comparison to the hyperlink graph, edges in the co-browsing graph have more homophily, such that a larger fraction of edges would be present in the lower deciles (corresponding to neighbors in the hyperlink graph that are more similar in leaning to the focal node).
Figure~\ref{fig:missing_edges}a shows that, indeed, as we move from left to right (i.e., as the leaning difference between neighbors in the hyperlink graph increases), the fraction of edges present in the co-browsing graph decreases.

The above analysis does not take into account the edge weights present in the co-browsing graph; i.e., it does not distinguish between the case where one single participant co-visited two domains from the case where a large number of participants did so.
To obtain the weighted version of Figure~\ref{fig:missing_edges}a, we computed, for each decile, the average weight of the corresponding edges in the co-browsing graph, with edges absent from the co-browsing graph contributing values of zero. (Note that the previous analysis is a special case of the present analysis, with all non-zero edge weights set to the value of~1.)
This way, we obtain, for each decile, the average number of users co-visiting the domains linked by the corresponding set of edges in the hyperlink graph.
The results, shown in Figure~\ref{fig:missing_edges}b, are consistent with those of the unweighted analysis (Figure~\ref{fig:missing_edges}a),
confirming that dissimilar (with respect to leaning) neighbors of a node are less sought out by users when browsing.

As mentioned above, a higher average degree could lead to a more diverse set of neighbors and hence, this result could be drastically different for nodes having vastly different degrees.
To investigate this possibility, we also repeated the above analysis after stratifying nodes into 10 equally sized group with respect to their degree in the hyperlink graph.
Inspecting version of Figure~\ref{fig:missing_edges} plotted separately for each of the 10 degree strata (not shown), we observed a similar pattern within each stratum, indicating the robustness of the result.

\xhdr{Multi-hop neighborhoods: random walks}
The above analysis only considered the immediate neighborhood of a node when assessing whether personal preferences are more polarized than the static link structure.
To understand how polarized the broader neighborhood of a node is, we conduct a random walk--based analysis, as follows.
Starting from each node $u$, we perform a random walk and, at each step $j$, measure the absolute difference $\delta_{uj}$ in leaning between $u$ and the node at step $j$.
Then, for each step $j$, we compute the \textit{neighborhood distance} $\delta_j$ at step $j$ as the mean value of $\delta_{uj}$ over all nodes $u$ in the graph.
To ensure the robustness of the results, we repeat the random walk 100 times for each node $u$ and use the mean $\delta_{uj}$ for each $u$ and $j$.
If a graph is more polarized, a node's neighborhood will be more similar to the node itself, and the neighborhood distance $\delta_j$ will be smaller.

We ran the above procedure (with random walks of length 10) on the following three graphs:
(i)~the co-browsing graph without edge weights,
(ii)~the co-browsing graph with edge weights,
and (iii)~the hyperlink graph.
Random walks in the different graphs capture different types of structure and participant choices:
The weighted co-browsing graph captures the participants' choices, biasing the random walk towards edges that were empirically taken by more participants. 
The unweighted co-browsing graph still takes participant choices into account to a certain extent, but makes transitions to all neighboring domains equally likely.
Finally, a random walk on the hyperlink graph simulates random browsing of the considered domains.

\begin{figure}[t]
    \centering
    \includegraphics[width=\columnwidth]{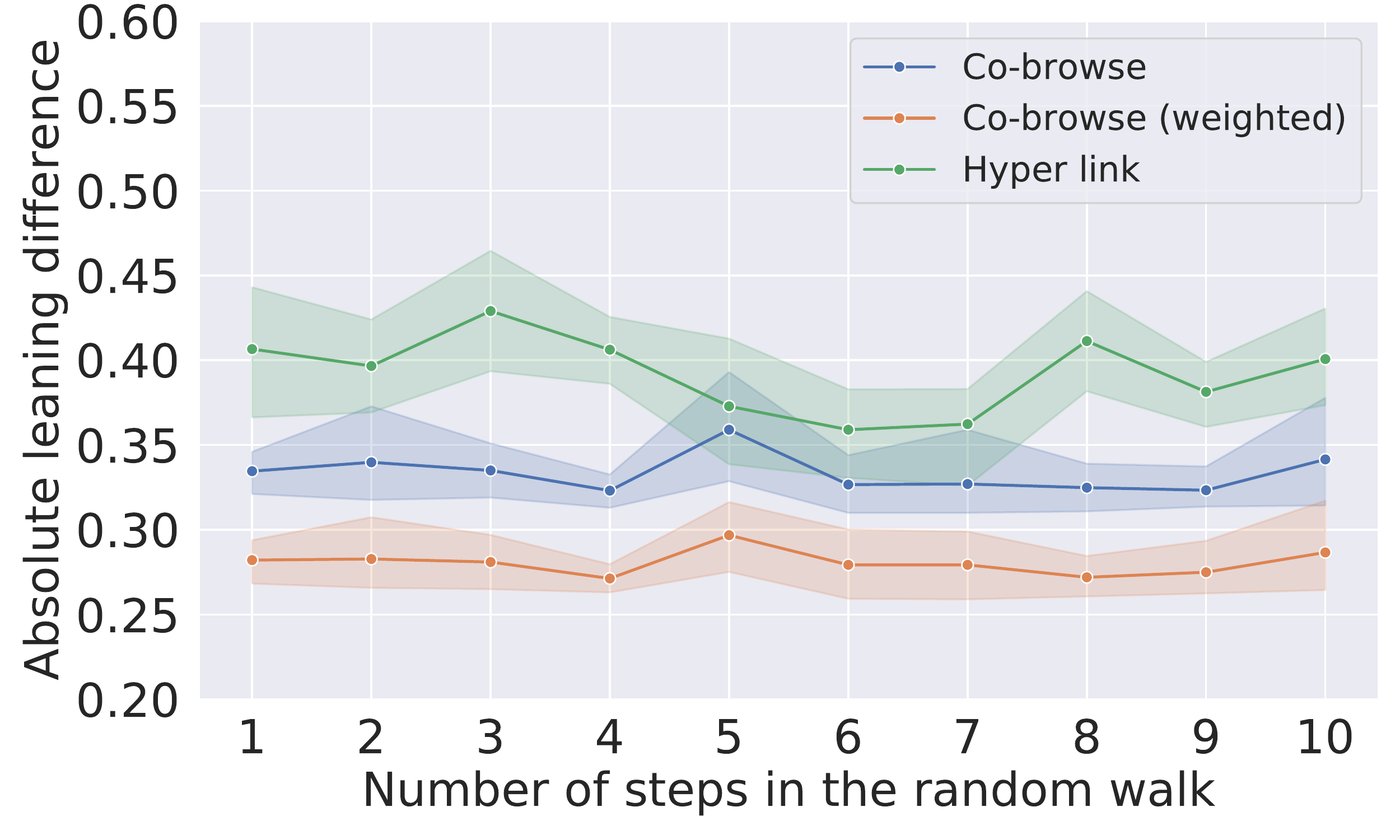}
    \caption{
    Leaning difference between nodes and their multi-hop neighborhoods in the weighted and unweighted co-browsing graphs and in the hyperlink graph, computed by performing random walks of length 10.
    Error bands show 95\% confidence intervals.
    The weighted co-browsing graph is most homophilic, followed by the unweighted co-browsing graph and the hyperlink graph.
    }
    \label{fig:randomwalk}
\vspace{-\baselineskip}
\end{figure}

The results are shown in Figure~\ref{fig:randomwalk}, with one curve for each of the three graphs.
Each point indicates the average neighborhood distance for a given step $j$ along the random walks.
%
%
%
The significantly lower values of neighborhood distance for the weighted co-browsing graph (red) indicate that participants' browsing choices are more biased towards domains with a similar leaning than what would be expected from someone randomly browsing the hyperlink graph induced by the included news domains (green).
There are also significant differences in the neighborhood distance values between the \textit{unweighted} co-browsing graph (blue) and the hyperlink graph (green), indicating inherently different (more homophilic) structure in the co-browsing graph, compared to the hyperlink graph, even after multiple hops of a random walking.
Figure~\ref{fig:randomwalk} thus strengthens the results of Figure~\ref{fig:missing_edges} beyond the immediate neighborhood of nodes. 

To summarize, the co-browsing graph appears to be significantly more polarized than the hyperlink graph, and we conclude that the polarization observed in the news consumption of the participants of our study cannot be explained by the hyperlink structure of the Web alone.
Rather, participants' explicit choices play an important role as well.

\section{Conclusions}

In this paper, we used browsing logs collected on the client side over the course of three weeks from a large opt-in panel of Firefox users in order to measure political polarization in online news consumption.
In various analyses, we provided evidence of pronounced polarization patterns in online news consumption.

By analyzing dwell time (the amount of time during which a user actively interacts with a webpage), we observed that users engage more deeply with news content aligned with their own leaning (estimated, by proxy, via the average leaning of the news they consume overall).

Applying a co-clustering algorithm to the bipartite browsing graph, which encodes average dwell times for all users on all visited domains, revealed three coherent, frequently co-visited groups of news domains with vastly different leaning distributions, corresponding to left-leaning, center, and right-leaning news domains, respectively---yet another indicator of polarization in news consumption.

Moving from the individual level to the population level, we contracted the bipartite user-by-domain browsing graph into a domain-by-domain co-browsing graph. Applying a community detection algorithm to the co-browsing graph gave rise to the same polarized pattern, with each cluster corresponding to one distinct political leaning.

A similarly polarized pattern---though less pronounced---was discovered by applying community detection to the hyperlink graph spanned by the news domains included in the study, rather than to the co-browsing graph.
Thus, in order to investigate the possibility that polarization in news consumption might be solely due to the inherent hyperlink structure of the Web, we compared the co-browsing and hyperlink graphs in various ways, concluding that the polarized hyperlink structure alone does not suffice to explain users' polarized news browsing behaviors.
Rather, users' explicit choices also contribute to the observed polarization.
Our findings are thus overall in line with previous studies on the role of user choices~\cite{bakshy2015exposure}, while contributing results based on a novel dataset of unprecedented detail and from novel angles of analysis.

\xhdr{Limitations}
Our work should be seen in the light of its limitations:
\begin{enumerate}
    \item Opt-in sample: While we strongly believe the opt-in nature of our recruitment is necessary for an ethical research approach, it may well have provided a biased sample. 
    \item Domain-level analysis: We aggregated all pages in a domain when constructing the browsing, co-browsing, and hyperlink graphs, essentially treating all pages in a domain as interchangeable. This was done in order to avoid sparsity issues that would arise when aggregating at the URL level instead.
    \item Personalization: Our analysis did not account for the role of personalization algorithms in browsing. While this is limiting to an extent, we point out that algorithmic personalization is still rare in online news~\cite{thurman2012future}.
    \item Correlation vs.\ causation:
    Although we showed that the polarization baked into the hyperlink structure of the Web alone does not suffice to explain users' polarized news consumption patterns, we do not claim to have revealed the causes of polarization in news consumption, which constitutes an important direction for future work.
\end{enumerate}

\xhdr{Future work: the role of the center}
Our analysis raises the profile of the role of the center as a clearly identifiable area in polarization studies, which proposes a potentially valuable role of centrist websites as a location for bipartisanship. These sites may themselves benefit from closer study and analysis, particularly when analyzed at the webpage or topic level. This becomes increasingly important when, as today, we see degrees of partisanship that are unprecedented and arguably unhealthy for society. Understanding where social worlds intersect provides opportunities for communication and potentially reconciliation within a polarized and separated populace.
We should ask: Can centrist websites serve as a ``demilitarized zone'' of the Web, and can an analysis such as ours, but focused on centrist websites, provide implications for the improved design of virtual spaces?

\section*{Acknowledgments}
Kiran Garimella was supported by a Michael Hammer Fellowship at MIT.
Robert West's lab was partly supported by grants from the Swiss National Science Foundation (grant 200021\_185043) and the Swiss Data Science Center, and by gifts from Google, Facebook, and Microsoft.

\balance
\bibliographystyle{aaai}
\bibliography{biblio}

\begin{thebibliography}{}

\bibitem[\protect\citeauthoryear{Bakshy, Messing, and
  Adamic}{2015}]{bakshy2015exposure}
Bakshy, E.; Messing, S.; and Adamic, L.~A.
\newblock 2015.
\newblock Exposure to ideologically diverse news and opinion on {F}acebook.
\newblock {\em Science} 348(6239):1130--1132.

\bibitem[\protect\citeauthoryear{Barber{\'a} \bgroup et al\mbox.\egroup
  }{2015}]{barbera2015tweeting}
Barber{\'a}, P.; Jost, J.~T.; Nagler, J.; Tucker, J.~A.; and Bonneau, R.
\newblock 2015.
\newblock Tweeting from left to right: Is online political communication more
  than an echo chamber?
\newblock {\em Psychological Science} 26(10):1531--1542.

\bibitem[\protect\citeauthoryear{Barthel \bgroup et al\mbox.\egroup
  }{2015}]{barthel2015evolving}
Barthel, M.; Shearer, E.; Gottfried, J.; and Mitchell, A.
\newblock 2015.
\newblock The evolving role of news on {Twitter and Facebook}.
\newblock {\em Pew Research Center}.

\bibitem[\protect\citeauthoryear{Batagelj and Zaversnik}{2003}]{batagelj2003m}
Batagelj, V., and Zaversnik, M.
\newblock 2003.
\newblock An {$O(m)$} algorithm for cores decomposition of networks.
\newblock {\em arXiv preprint cs/0310049}.

\bibitem[\protect\citeauthoryear{De~Meo \bgroup et al\mbox.\egroup
  }{2011}]{de2011generalized}
De~Meo, P.; Ferrara, E.; Fiumara, G.; and Provetti, A.
\newblock 2011.
\newblock Generalized {L}ouvain method for community detection in large
  networks.
\newblock In {\em ICISDA}.

\bibitem[\protect\citeauthoryear{Dhillon}{2001}]{dhillon2001co}
Dhillon, I.~S.
\newblock 2001.
\newblock Co-clustering documents and words using bipartite spectral graph
  partitioning.
\newblock In {\em KDD}.

\bibitem[\protect\citeauthoryear{Flaxman, Goel, and
  Rao}{2016}]{flaxman2016filter}
Flaxman, S.; Goel, S.; and Rao, J.~M.
\newblock 2016.
\newblock Filter bubbles, echo chambers, and online news consumption.
\newblock {\em Public Opinion Quarterly} 80(S1):298--320.

\bibitem[\protect\citeauthoryear{Garimella \bgroup et al\mbox.\egroup
  }{2018}]{garimella2018political}
Garimella, K.; De~Francisci~Morales, G.; Gionis, A.; and Mathioudakis, M.
\newblock 2018.
\newblock Political discourse on social media: Echo chambers, gatekeepers, and
  the price of bipartisanship.
\newblock In {\em WWW}.

\bibitem[\protect\citeauthoryear{Garrett}{2009}]{garrett2009echo}
Garrett, R.~K.
\newblock 2009.
\newblock Echo chambers online? politically motivated selective exposure among
  {I}nternet news users.
\newblock {\em Journal of Computer-Mediated Communication} 14(2):265--285.

\bibitem[\protect\citeauthoryear{Geiger}{2019}]{geiger2019key}
Geiger, A.
\newblock 2019.
\newblock Key findings about the online news landscape in {A}merica.
\newblock {\em Pew Research Center}.

\bibitem[\protect\citeauthoryear{Gentzkow and
  Shapiro}{2011}]{gentzkow2011ideological}
Gentzkow, M., and Shapiro, J.~M.
\newblock 2011.
\newblock Ideological segregation online and offline.
\newblock {\em The Quarterly Journal of Economics} 126(4):1799--1839.

\bibitem[\protect\citeauthoryear{Guess}{2018}]{guess2016media}
Guess, A.~M.
\newblock 2018.
\newblock ({Almost}) everything in moderation: New evidence on {A}mericans'
  online media diets.
\newblock {\em American Journal of Political Science}.

\bibitem[\protect\citeauthoryear{Iyengar and Hahn}{2009}]{iyengar2009red}
Iyengar, S., and Hahn, K.~S.
\newblock 2009.
\newblock Red media, blue media: Evidence of ideological selectivity in media
  use.
\newblock {\em Journal of Communication} 59(1):19--39.

\bibitem[\protect\citeauthoryear{Iyengar \bgroup et al\mbox.\egroup
  }{2019}]{iyengar2019origins}
Iyengar, S.; Lelkes, Y.; Levendusky, M.; Malhotra, N.; and Westwood, S.~J.
\newblock 2019.
\newblock The origins and consequences of affective polarization in the {United
  States}.
\newblock {\em Annual Review of Political Science} 22:129--146.

\bibitem[\protect\citeauthoryear{Klapper}{1960}]{klapper1960effects}
Klapper, J.~T.
\newblock 1960.
\newblock {\em The effects of mass communication}.
\newblock Free Press.

\bibitem[\protect\citeauthoryear{Kulshrestha \bgroup et al\mbox.\egroup
  }{2017}]{kulshrestha2017quantifying}
Kulshrestha, J.; Eslami, M.; Messias, J.; Zafar, M.~B.; Ghosh, S.; Gummadi,
  K.~P.; and Karahalios, K.
\newblock 2017.
\newblock Quantifying search bias: Investigating sources of bias for political
  searches in social media.
\newblock In {\em CSCW}.

\bibitem[\protect\citeauthoryear{Lahoti, Garimella, and
  Gionis}{2018}]{lahoti2018joint}
Lahoti, P.; Garimella, K.; and Gionis, A.
\newblock 2018.
\newblock Joint non-negative matrix factorization for learning ideological
  leaning on {Twitter}.
\newblock In {\em WSDM}.

\bibitem[\protect\citeauthoryear{Messing and
  Westwood}{2014}]{messing2014selective}
Messing, S., and Westwood, S.~J.
\newblock 2014.
\newblock Selective exposure in the age of social media: Endorsements trump
  partisan source affiliation when selecting news online.
\newblock {\em Communication Research} 41(8):1042--1063.

\bibitem[\protect\citeauthoryear{Meusel \bgroup et al\mbox.\egroup
  }{2014}]{meusel2014graph}
Meusel, R.; Vigna, S.; Lehmberg, O.; and Bizer, C.
\newblock 2014.
\newblock Graph structure in the {Web}---revisited: A trick of the heavy tail.
\newblock In {\em WWW}.

\bibitem[\protect\citeauthoryear{Narayanan \bgroup et al\mbox.\egroup
  }{2018}]{narayanan2018polarization}
Narayanan, V.; Barash, V.; Kelly, J.; Kollanyi, B.; Neudert, L.-M.; and Howard,
  P.~N.
\newblock 2018.
\newblock Polarization, partisanship and junk news consumption over social
  media in the {US}.
\newblock {\em arXiv preprint arXiv:1803.01845}.

\bibitem[\protect\citeauthoryear{Nelson and Webster}{2017}]{nelson2017myth}
Nelson, J.~L., and Webster, J.~G.
\newblock 2017.
\newblock The myth of partisan selective exposure: A portrait of the online
  political news audience.
\newblock {\em Social Media + Society} 3(3).

\bibitem[\protect\citeauthoryear{Nikolov \bgroup et al\mbox.\egroup
  }{2019}]{nikolov2019quantifying}
Nikolov, D.; Lalmas, M.; Flammini, A.; and Menczer, F.
\newblock 2019.
\newblock Quantifying biases in online information exposure.
\newblock {\em JAIST} 70(3):218--229.

\bibitem[\protect\citeauthoryear{Pariser}{2011}]{pariser2011filter}
Pariser, E.
\newblock 2011.
\newblock {\em The filter bubble: What the Internet is hiding from you}.
\newblock Penguin UK.

\bibitem[\protect\citeauthoryear{Parse.ly}{2016}]{parsley}
Parse.ly.
\newblock 2016.
\newblock Network referrer dashboard.
\newblock Data retrieved from
  \url{https://www.parse.ly/resources/data-studies/referrer-dashboard}.

\bibitem[\protect\citeauthoryear{Peterson, Goel, and
  Iyengar}{2018}]{peterson2018echo}
Peterson, E.; Goel, S.; and Iyengar, S.
\newblock 2018.
\newblock Echo chambers and partisan polarization: Evidence from the 2016
  presidential campaign.
\newblock {\em Political Science Research and Methods}.

\bibitem[\protect\citeauthoryear{Raghavan, Anderson, and
  Kleinberg}{2018}]{raghavan2018mapping}
Raghavan, M.; Anderson, A.; and Kleinberg, J.
\newblock 2018.
\newblock Mapping the invocation structure of online political interaction.
\newblock In {\em WWW}.

\bibitem[\protect\citeauthoryear{Ribeiro \bgroup et al\mbox.\egroup
  }{2018}]{ribeiro2018media}
Ribeiro, F.~N.; Henrique, L.; Benevenuto, F.; Chakraborty, A.; Kulshrestha, J.;
  Babaei, M.; and Gummadi, K.~P.
\newblock 2018.
\newblock Media bias monitor: Quantifying biases of social media news outlets
  at large-scale.
\newblock In {\em ICWSM}.

\bibitem[\protect\citeauthoryear{Robertson \bgroup et al\mbox.\egroup
  }{2018}]{robertson2018auditing}
Robertson, R.~E.; Jiang, S.; Joseph, K.; Friedland, L.; Lazer, D.; and Wilson,
  C.
\newblock 2018.
\newblock Auditing partisan audience bias within {G}oogle search.
\newblock {\em Proceedings of the ACM on Human-Computer Interaction (CSCW)}
  2:1--22.

\bibitem[\protect\citeauthoryear{Rogowski and
  Sutherland}{2016}]{rogowski2016ideology}
Rogowski, J.~C., and Sutherland, J.~L.
\newblock 2016.
\newblock How ideology fuels affective polarization.
\newblock {\em Political Behavior} 38(2).

\bibitem[\protect\citeauthoryear{Role, Morbieu, and
  Nadif}{2018}]{role2018coclust}
Role, F.; Morbieu, S.; and Nadif, M.
\newblock 2018.
\newblock {CoClust}: A {P}ython package for co-clustering.

\bibitem[\protect\citeauthoryear{Schmidt \bgroup et al\mbox.\egroup
  }{2017}]{schmidt2017anatomy}
Schmidt, A.~L.; Zollo, F.; Del~Vicario, M.; Bessi, A.; Scala, A.; Caldarelli,
  G.; Stanley, H.~E.; and Quattrociocchi, W.
\newblock 2017.
\newblock Anatomy of news consumption on {F}acebook.
\newblock {\em Proceedings of the National Academy of Sciences}
  114(12):3035--3039.

\bibitem[\protect\citeauthoryear{Sears and Freedman}{1967}]{sears1967selective}
Sears, D.~O., and Freedman, J.~L.
\newblock 1967.
\newblock Selective exposure to information: A critical review.
\newblock {\em Public Opinion Quarterly} 31(2):194--213.

\bibitem[\protect\citeauthoryear{Stroud}{2010}]{stroud2010polarization}
Stroud, N.~J.
\newblock 2010.
\newblock Polarization and partisan selective exposure.
\newblock {\em Journal of Communication} 60(3):556--576.

\bibitem[\protect\citeauthoryear{Sunstein}{2009}]{sunstein2009republic}
Sunstein, C.~R.
\newblock 2009.
\newblock {\em Republic.com 2.0}.
\newblock Princeton University Press.

\bibitem[\protect\citeauthoryear{Thurman and
  Schifferes}{2012}]{thurman2012future}
Thurman, N., and Schifferes, S.
\newblock 2012.
\newblock The future of personalization at news websites: Lessons from a
  longitudinal study.
\newblock {\em Journalism Studies} 13(5-6):775--790.

\bibitem[\protect\citeauthoryear{Webster and
  Abramowitz}{2017}]{webster2017ideological}
Webster, S.~W., and Abramowitz, A.~I.
\newblock 2017.
\newblock The ideological foundations of affective polarization in the {US}
  electorate.
\newblock {\em American Politics Research} 45(4):621--647.

\end{thebibliography}

\end{document}